\documentclass[12pt]{article}

\usepackage{latexsym}
\usepackage{amsthm}
\usepackage{amsmath}
\usepackage{graphicx}   
\usepackage{amssymb}

\begin{document}

\begin{center}
\vspace{1cm}

{\Large \bf Split Supersymmetry and Dark Matter Generation}

\vspace{0.8cm}

{\bf Jared Kaplan\footnote{email: kaplan@physics.harvard.edu}}

\vspace{.5cm}

{\it Jefferson Laboratory of Physics, Harvard University,\\
         Cambridge, Massachusetts 02138, USA}

\end{center}
\vspace{1cm}

\begin{abstract}
\medskip

We analyze a simple Split Supersymmetry scenario where fermion masses come from
anomaly mediation, yielding $m_s \sim 1000$ TeV, $m_{3/2} \sim 100$ TeV, and
$m_f \sim 1 $ TeV.  We consider non-thermal dark matter production in the presence of
moduli, and we find that the decay chains of moduli $\to$ LSPs and 
moduli $\to$ gravitinos $\to$ LSPs generate dark matter more efficiently than 
perturbative freeze-out, allowing for a light, LHC visible spectrum.
These decaying moduli can also weaken cosmological constraints on the axion decay constant.
With squark masses of order $1000$ TeV, LHC gluinos will decay millimeters from their 
primary vertices, resulting in a striking experimental signature, and the suppression of Flavor 
Changing Neutral Currents is almost sufficient to allow arbitrary mixing in squark mass matrices.

\end{abstract}

\bigskip

\section{Introduction}

With the Large Hadron Collider only a few years away, a new division has arisen
in the particle physics community -- is the unnaturalness of the standard model 
a problem to be solved through clever model building, or is it a hint
that physics at the TeV scale is different than we have imagined, so
that the values of dimensionful parameters are determined by anthropic fine tuning?
As physicists, we would prefer a `physical' explanation of small numbers 
such as the cosmological constant and the higgs mass, but the success of 
Weinberg's prediction of the cosmological constant \cite{wein} and the level of fine tuning 
necessary even in our best theories forces us to take the anthropic argument 
seriously.  We will be taking it seriously for the present work.

Yet the mind of an anthropically-motivated model builder is a troubled one, for
it seems that this profound shift in our worldview has had the unfortunate side effect of 
putting our whole enterprise out of business.  If the weak scale is determined by anthropic
selection, perhaps there is no new physics at the TeV scale, and the LHC and even the ILC
will be colossal disappointments.

There are two simple reasons to remain hopeful about new TeV scale physics:  
gauge coupling unification and dark matter.  The former requires new particles charged
under the standard model gauge group, and the latter requires a new stable particle
with the correct relic density.  If we assume for maximal simplicity that these two problems
have a common solution, then it is reasonable to assume that our dark matter candidate will have a
weak interaction cross section.  Furthermore, if we assume that the relic dark matter density
is determined by perturbative freeze-out, then we can expect new physics at the TeV scale,
tightly constrained by gauge coupling unification.  Some of these assumptions may be wrong,
but that is not necessarily a bad thing -- the minimal model that follows from this reasoning
will probably be invisible to the LHC \cite{mindmunif}.

A very popular set of models that result from this methodology fall under the 
heading of Split Supersymmetry \cite{split}, where the scalar superpartners necessitated 
by supersymmetry are very heavy, while the fermions, protected by chiral symmetry, 
lie near the TeV scale.  The purpose of the present work is to show that in one particularly
simple model of Split Supersymmetry, perturbative freeze-out is not the dominant
mechanism for generating dark matter.  However, the mechanism that will replace it
is more efficient, allowing for an even lighter, more LHC-visible spectrum.  This seems like
good news, because in standard Split Supersymmetry scenarios \cite{aspsplit}, \cite{wells}, 
dark matter is expected to be too heavy to be seen at the LHC; however we will see that
in our case there is a danger of overclosure.  One of the most elegant scenarios is ruled out,
but we explore several ideas that can rescue it, and the scenario of Moroi and Randall 
\cite{modtosm} remains a viable possibility.

To understand our mechanism, we first need to explain the spectrum of the model.  As shown in 
\cite{anom}, \cite{outworld}, wherever supersymmetry is broken, there will be 
visible sector supersymmetry breaking effects from Anomaly Mediation.  Furthermore, 
the methodology of effective field theory requires that we include in the Lagrangian all higher dimension 
operators allowed by symmetry, with appropriate suppression by inverse powers of the cutoff.  
When supersymmetry is broken in a hidden sector, F and D term VEVs in these operators will 
induce supersymmetry breaking in the visible sector.  These two effects are completely generic, requiring no
theoretical gymnastics, so a model where supersymmetry breaking arises only in this way
would be particularly elegant.  This is the model that we will analyze; it was studied
for related reasons in \cite{wells}.

With Anomaly Mediated Supersymmetry Breaking, gaugino masses are given by a loop 
suppression factor times $m_{3/2}$, so we expect that $m_{3/2}  \sim 100$ TeV so that 
new fermions are near a TeV.  Contributions from higher dimension operators suggest 
that the scalar mass scale $m_s \sim 1000$ TeV.  In supersymmetric theories, there generically exist weakly interacting 
moduli that get masses of order $m_s$ after supersymmetry breaking.  Now we
see the complexity of the cosmology -- not only do we have to account for the 
perturbative freeze out of the LSPs, but we also need to consider processes involving
late-decaying moduli and gravitinos.  These late decaying particles produce a great deal of 
entropy, potentially weakening cosmological constraints on the properties of axions.  

We will see that with our mass spectrum, dark matter production is dominated by two
potential decay chains: moduli $\to$ LSPs and moduli $\to$ gravitinos $\to$ LSPs.  
Only the first process is available for $m_\phi < 2m_{3/2}$,\footnote{Of course this depends on 
the available gravitino-producing modulus decay channels -- the strict limit is $m_\phi < m_{3/2}$ -- 
but under reasonable assumptions $\phi \to 2 \psi_{3/2}$ is the only available decay mode.} 
whereas both decay chains are
open for $m_\phi > 2m_{3/2}$ -- thus the physics is qualitatively different for different values
of the modulus mass.  The first process was studied in \cite{modtosm}, and it naturally leads to the correct relic
density in our scenario when $m_\phi \lesssim m_{3/2}$.  The most naive version of the second
decay chain is too efficient \cite{newdecay}, \cite{newdecay2}, \cite{dine}, so we consider 
a variety of mechanisms to alleviate this problem,
including symmetries in the modulus sector, scenarios with multiple moduli, KKLT type moduli,
and supersymmetry breaking dominated by D-terms.  Of course we do not need these mechanisms
for a successful cosmology if $m_\phi \lesssim m_{3/2}$, but it is interesting to explore all 
reasonable options.

Our model also has an exciting LHC signature.  The gluino must decay through a virtual 
squark, so with $1000$ TeV squarks, it is very likely that the LHC will 
see displaced gluino vertices if gluino production is kinematically allowed.  In fact, we 
expect that gluinos will be copiously produced, since cosmology suggests a light spectrum.

We are also in an interesting region for Flavor Changing Neutral Currents.  
The strongest constraints from FCNC come from the $\epsilon$ 
parameter of the K-$\bar{\mathrm{K}}$ system, which depends on the imaginary
parts of the mass insertion parameters.  Setting all mass insertions equal to a common 
value $\delta$ (there are no significant cancellations), we find from \cite{FCNC} that
\begin{equation}
\epsilon = 3 \times 10^{-3} \delta \left(\frac{1000 \ \mathrm{TeV}}{m_s}\right)^2
\end{equation}
Experiments constrain $\epsilon < 2 \times 10^{-3}$, so for $m_s \approx 1000$ TeV,
we find that as long as $\delta \lesssim 1/2$, our scalar mass scale does not conflict with observations
of FCNC.  We also expect that electric dipole moments induced by new interactions may be visible at next 
generation experiments, as shown in \cite{aspsplit}, and our spectrum may help to explain neutrino masses, 
as examined in \cite{wells}.

The outline of this paper is as follows.  In section two we display the mass spectrum
of our model.  In section three we consider the cosmological implications of moduli, gravitinos, and LSP
(Wino) dark matter.  In section four we show that displaced gluino vertices will be a generic LHC 
signature for our model, and with section five we conclude.

\section{The Mass Spectrum}

We obtain the fermion mass spectrum
\begin{eqnarray}
m_{3/2} & = & \frac{\langle W \rangle}{M_p^2} = \langle F_\phi \rangle \sim 100 \ \mathrm{TeV} \\
m_{\tilde{g}} & = & \frac{\beta(g)}{2g} \langle F_\phi \rangle \sim 1 \ \mathrm{TeV}
\end{eqnarray}
from anomaly mediated supersymmetry breaking \cite{anom}, \cite{outworld}.  The details of hidden 
sector supersymmetry breaking are unimportant because the hidden sector only communicates 
with standard model particles through the auxilary field of supergravity and through higher dimension
operators in the Kahler potential.  We assume 
that the scalars acquire masses from operators such as
\begin{equation}
L \supset \int d^4 \theta \frac{1}{M_{GUT}^2} X^{\dag}X Q^{\dag} Q
\end{equation}
where $X$ is a hidden sector field, because these terms cannot be forbidden by any symmetry.  
We expect these operators to arise generically when we integrate out 
GUT or string scale particles, making them a factor of $10$ - $100$ times
larger than $m_{3/2}$.  Thus we can estimate that
\begin{equation}
m_s = m_{3/2} \frac{M_{pl}}{M_{GUT}} \sim 1000 \ \mathrm{TeV} ,
\end{equation}
completing our rough picture of the mass spectrum.

It is also important to consider the generation and the effects of the $\mu$ and $B\mu$ terms.
Will $\mu$ and $B \mu$ be near $m_s$, $m_{3/2}$, or at the TeV scale?  First, there 
can be contributions from R-symmetry breaking spurionic operators 
$X = 1 + \theta^2 m_s$ such as
\begin{eqnarray}
L & \supset & \int d^4 \theta \epsilon X^{\dag} X H_1 H_2 \\
L & \supset & \int d^4 \theta \epsilon X^{\dag} H_1 H_2
\end{eqnarray}
where the factor of $\epsilon$ is included to parameterize a possible approximate
PQ symmetry.  This gives rise to $B \mu \sim \epsilon m_s^2$ and $\mu \sim \epsilon m_s$.  
In most split supersymmetry scenarios we use R-symmetry to prohibit 
such spurionic contributions because they produce a nearly degenerate spectrum.  However,
in our scenario $m_{3/2}/m_f \sim 100$, which can be conveniently explained as
a loop factor from anomaly mediation.  Thus such operators may be
permitted.  If supersymmetry breaking does not lead to R-symmetry breaking 
we only have spurions\footnote{As explained in \cite{aspsplit}, this does not 
require that supersymmetry breaking comes from gauge superfields, but only that 
supersymmetry breaking is not accompanied by R-symmetry breaking.} such as
\begin{equation}
L \supset \int d^4 \theta \epsilon Y H_1 H_2
\end{equation}
with $Y = 1 + \theta^4 m_s^2$, which contribute to $B \mu$ only.
Finally, there are contributions from the conformal compensator of supergravity
\begin{equation}
L \supset \int d^4 \theta \epsilon \phi^{\dag} \phi H_1 H_2 
\end{equation}
where $\phi = 1 + \theta^2 m_{3/2}$, so that $\mu \sim \epsilon m_{3/2}$.  
Thus we find that $B \mu \approx \epsilon m_s^2$ in all cases,
and either $\mu \approx \epsilon m_s$ or $\mu \approx \epsilon m_{3/2}$ .  We
can take $\epsilon \approx 1/100$ to explain the top-bottom mass hierarchy,
giving a $\mu$ term at the TeV scale.

The presence of the $\mu$ term modifies the gaugino masses \cite{anom}, \cite{outworld} so that
\begin{eqnarray} \label{spectrum}
m_{\tilde{b}} & = & 8.9 \times 10^{-3} \left(1 - \frac{f(\mu^2/m_A^2)}{11} \right) m_{3/2} \\
m_{\tilde{w}} & = & 2.7 \times 10^{-3} \left(1 - f(\mu^2/m_A^2) \right) m_{3/2} \\
m_{\tilde{g}} & = & 2.6 \times 10^{-2} m_{3/2}
\end{eqnarray}
where $m_A^2 \sin(2 \beta) = B \mu$ and the function
\begin{equation}
f(x) = \frac{2 x \ln(x)}{x - 1}
\end{equation}
For small values of $x$ (the expected case) this is a small effect.  It increases the separation 
between the wino, bino, and gluino mass scales, but there is no real qualitative change in the spectrum.  
For large $x$ we can obtain a very light Wino LSP, although we would 
need to abandon the PQ symmetry; this will be breifly considered in section 3.3.3.

\section{Cosmology}

Our variant of the Split Supersymmetry mass spectrum includes $100$-$10,000$ TeV moduli
field(s) $\phi$ and a $100$ TeV gravitino, so we must check that these new ingredients do not 
disturb Big Bang Nucleosynthesis, and that the correct relic abundance of LSP 
dark matter candidates obtains.  We begin with some general issues about our universe's
history, and then we consider the possibility that our moduli fields weaken cosmological bounds on
the axion.  Finally, we explain the details of our proposed LSP generation mechanism.

We can think of our modulus field as though it were an inflaton -- it begins with a VEV $\phi_0$
that can be of order $M_{pl}$.  Once the hubble constant decreases to $H \sim m_\phi$, 
the modulus begins to oscillate and its energy density red-shifts like that of matter.  This occurs very
early in the universe's history, with $T_{roll} \sim \sqrt{M_{pl} m_{\phi}} \sim 10^9$ TeV.  At a time 
$t_{eq}$ shortly after $\phi$ begins to roll (assuming $\phi$ begins with a Planckian VEV), 
$\rho_\phi = \rho_{Rad}$, and from this point until $t_{decay} = \Gamma_{tot}^{-1}$ 
the universe is modulus dominated.  We take the modulus decay rate as\footnote{There is certainly
a significant uncertainty because we do not know any of the $O(1)$ couplings involved.  However,
we do know that modulus decays to matter in the complex representation of a symmetry group 
\cite{modtosm}, \cite{newdecay}, \cite{dine} receive a suppression proportional to a power of $m_{mat}/m_\phi$, 
which means that moduli decay dominantly to gauge bosons, gauginos, and possibly the heavy scalars.}
\begin{equation}
\Gamma_{tot} = \frac{N}{16 \pi} \frac{m_\phi^3}{M_{pl}^2} = 
\left(10^{-5} \ \mathrm{sec}\right)^{-1} \left( \frac{N}{10} \right) \left(\frac{m_\phi}{1000 \ \mathrm{TeV}}\right)^3
\end{equation}
where $M_{pl} = 2.4 \times 10^{18}$ GeV is the reduced Planck mass.  
We can think of $N$ as the number of available light decay modes, although it is strongly dependent
on unknown, $O(1)$ couplings.  We see that the modulus decays well before BBN. 
The decaying modulus will dilute particles left over from the original inflaton decay, 
reheat the universe, and re-populate the universe with gravitinos.

Any thermal gravitinos produced during a prior inflationary reheating are diluted by the 
entropy from the modulus decay, which is given by
\begin{equation}
\frac{s_{after}}{s_{before}} = \left(\frac{\rho_\phi(t_{decay})}{\rho_R(t_{decay})}\right)^{3/4} 
\sim \frac{M_{pl}}{m_\phi} \left(\frac{\phi_0}{M_{pl}}\right)^2
\end{equation}
where the ratio of the energy densities at the time of modulus decay simply comes from
the two different equations of state.  If we assume $\phi_0$ is in the neighborhood 
of $M_{pl}$, then the entropy released is very large.  
Modulus decay reheats the universe to a temperature \cite{modtogravmod} 
\begin{equation}
T_R \approx 1.1 g_*^{-1/4} \sqrt{\Gamma_{tot} M_{pl}}
= 180 \ \mathrm{MeV} \times \left( \frac{N}{10} \right)^{1/2} \left(\frac{g_*}{10.75}\right)^{-1/4} \left(\frac{m_\phi}{1000 \ \mathrm{TeV}}\right)^{3/2} 
\end{equation}
which is very low, so we certainly do not thermally regenerate heavy particles.  To avoid
disturbing BBN, we need $T_R \gg 1$ MeV, so $m_\phi \gg 30$ TeV.
Similarly, the gravitino lifetime is \cite{moroithesis}
\begin{equation}
\tau_{3/2} = 5 \times 10^{-2} \ \mathrm{sec} \left(\frac{100 \ \mathrm{TeV}}{m_{3/2}}\right)^3
\end{equation}
so for $m_{3/2} > 60$ TeV gravitinos do not disturb BBN \cite{BBNgrav}, \cite{newdecay} 
even if they are re-introduced by modulus decay.  Note, however, that the lifetime is increased
in split supersymmetry because the gravitino cannot decay to states involving the heavy scalars.

Dark matter is produced in two different ways: directly from decaying moduli, and indirectly 
from decaying gravitinos.  Thermal relic gravitinos are diluted to negligible levels by modulus decay, so gravitinos
can only arise from the decay of heavy moduli with $m_\phi > 2m_{3/2}$.  Thus there are two qualitatively
different types of moduli -- those with $m_\phi > 2 m_{3/2}$ and those with $m_\phi < 2 m_{3/2}$.
In the following subsection we consider direct modulus-LSP interactions, focusing on moduli with $m_\phi < 2 m_{3/2}$,
and commenting on heavier moduli.  Then in the next subsection we consider gravitino production from moduli 
with $m_\phi > 2 m_{3/2}$.  It turns out that dark matter production from gravitinos is more efficient than
direct production from modulus decay; this is essentially because the dark matter from gravitinos is produced
later in the universe's history, so it accounts for a larger fraction of the universe's energy density.

\subsection{Direct Modulus Decay to Dark Matter}

We expect that the modulus will have generic string or Planck suppressed couplings to
MSSM fields.  For example, we may have interactions such as
\begin{equation}
\int d^2\theta \frac{\Phi}{M_{pl}} W_\alpha W^\alpha  \ \ \ \implies \ \ \
L \supset  \frac{\phi}{M_{pl}} F_{\mu \nu}^2 + \frac{F_\phi}{M_{pl}} \lambda \lambda ,
\end{equation}
where $\lambda$ may be any gaugino\footnote{Note that if $\Phi$ acquires
a large $F$-component VEV, then this operator could give dangerously large gaugino masses, so we must 
assume that either $F_\phi$ is small or that the coefficient of this operator is suppressed.}.  
If $\Phi$ has a supersymmetric mass, then $F_\phi \propto m_\phi \phi$, and both terms
give large decay rates, but in other situations where $m_\phi$ only arises after supersymmetry 
breaking, we may have $\frac{d F_\phi}{d \phi} < m_\phi$ and the fermionic decay mode will
be suppressed by the fermion mass.  
Due to R-Parity conservation, this implies that an order one fraction
of the `modulus particles' may eventually produce an LSP.  The number density of `modulus
particles' relative to the entropy just after reheating is roughly
\begin{equation}
Y_\phi  = \left( \frac{m_{\phi}}{T_R} \right)^{-1} \sim \left( \frac{m_{\phi}}{M_{Pl}} \right)^{1/2}
\end{equation}
If this were the whole story, we would overclose the universe.  If the branching ratio to fermions 
is suppressed by $m_f^2/m_\phi^2$ an acceptable dark matter density might be possible.

However, the LSPs produced by modulus decay pair annihilate \cite{modtosm} until their annihilation
rate is less than $H$, after which they freeze out.  Although the Wino's produced by the decaying modulus
are ultra-relativistic, they acquire a thermal distribution with temperature $T_R$ from interactions with the
radiation bath.  For instance, the process $\tilde{W^0} \nu \to \tilde{W^0} \nu$ proceeds more quickly 
than $H$ for $T_R \gg 1$ MeV, rendering the Winos non-relativistic.  The 
annihilation cross section for Wino dark matter \cite{modtosm} into $W^{\pm}$ pairs is 
\begin{equation}
\langle v_{rel} \sigma \rangle = \frac{g^2}{2 \pi} \frac{1}{m_{LSP}^2} \frac{(1-x_W)^{3/2}}{(2-x_W)^2}
\end{equation}
in the non-relativistic limit, with $x_W = m_W^2/m_{LSP}^2$.  Thus the maximal dark matter relic density 
is reduced to
\begin{equation} \label{annih}
n_{LSP} \sim \left. \frac{3 H}{2 \langle v \sigma \rangle} \right|_{T_R} \sim \frac{T_R^2}{M_{pl} \langle v \sigma \rangle} .
\end{equation}
Note that this will apply to any late decay that produces dark matter, including gravitino decay.
Thus we find \cite{modtosm}
\begin{equation}
\frac{\rho}{s} \approx 4 \times 10^{-10} \ \mathrm{GeV} \times \left( \frac{10}{N} \right) 
\left(\frac{g_*}{10.75} \right)^{-1/4} \frac{(2-x_W)^2}{(1-x_W)^{3/2}} 
\left(\frac{m_{LSP}}{100  \ \mathrm{GeV}} \right)^3 \left(\frac{100  \ \mathrm{TeV}}{m_\phi} \right)^{3/2}
\end{equation}
where we have taken $g_* = 10.75$.  We need $\rho/s \approx 4 \times 10^{-10}$ GeV \cite{WMAP} 
to account for dark matter, so if there is a single light modulus
with $m_\phi < 2m_{3/2}$, we can easily obtain the correct relic abundance \cite{modtosm}.  
Since the decay of this modulus will not produce gravitinos, we have a viable cosmology.

If $m_\phi \gg m_{3/2}$ then after modulus decay we are left with a negligable density of LSPs, 
so next we consider the consequences of gravitino production in scenarios with such a heavy modulus.

\subsection{Moduli-Gravitino Interactions and Indirect Dark Matter Production}

There are two potential channels in which modulus decay produces gravitinos,
\begin{eqnarray}
\phi & \to & \tilde{\phi} + \psi_{3/2} \\
\phi & \to & 2\psi_{3/2}
\end{eqnarray}
where $\tilde{\phi}$ is the modulino, the superpartner of the modulus field.
We do not expect the modulus to decay to a gravitino and another fermion
because the other fermion would reside in the hidden sector, and we expect that 
hidden sector fields are very massive.  For now we will assume that the modulino is
very heavy, so that the first channel is also forbidden by kinematics -- later we will see that
relaxing this assumption does not help us to obtain a viable cosmology.  In this section
we consider the limit that $m_\phi \gg m_{3/2}$, and we begin with the two gravitino channel.

Gravitinos interact primarily through their longitudinal (helicity $\pm 1/2$) components,
which correspond to the goldstino, the field that is `eaten' in the superhiggs mechanism.
Thus the goldstone boson equivalence theorem becomes the goldstino equivalence theorem,
which simply tells us that amplitudes involving longitudinal gravitinos can be computed
using the goldstino instead.
In the appendix it is shown that the only two derivative operator involving a scalar and 
two non-linearly realized, derivatively coupled goldstinos (longitudinal gravitinos) is
\begin{equation}
\frac{1}{F^\dag} \partial_\nu s_\chi^\dag \chi \sigma^\nu \bar{\sigma}^\mu \partial_\mu \chi 
- \frac{1}{F} \partial_\nu s_\chi \chi^\dag \bar{\sigma}^\mu \sigma^\nu \partial_\mu chi^\dag .
\end{equation}
where the `sgoldstino' field is
\begin{equation}
s_\chi = \sum_i \frac{F_i}{F} \phi_i ,
\end{equation}
and $F^2 = \Sigma_i F_i^2 + \Sigma_a D_a^2$ gives the goldstino decay constant.
The sgoldstino is generally not a mass eigenstate -- it is a blend of many different
scalar fields.  Thus the rate for moduli decay to goldstino pairs is determined entirely
by the overlap of the mass eigenstate modulus with the sgoldstino.

In supergravity one expects higher dimension operators to be present in
the Kahler potential, with suppression of order the GUT, string, or Planck scale.
These operators tend to shift the VEVs of scalar fields and auxiliary fields.
We would expect the effect on a modulus to be proportional to $1/m_{\phi}$
because the mass tends to stabilize the field, so on dimensional grounds we 
can estimate that
\begin{equation}
F_{\Phi} \sim \frac{\mu_{susy}^4}{m_{\phi} M_{pl}} \sim \frac{m_{3/2}}{m_\phi}F
\end{equation}
for the auxiliary field of a heavy modulus.  A more rigorous argument for this conclusion
was given in \cite{newdecay2}, and it seems to be generically true in specific models
in the absence of special symmetries\footnote{It continues to be true when the modulus has
a shift symmetry, although as we will discuss later there are exceptions in some models,
such as specific versions of KKLT \cite{dine}.}.  Assuming that the $\phi$ field really is a mass
eigenstate, this implies a rate for $\phi \to 2 \psi_{3/2}$
\begin{equation}
\Gamma_{3/2} = \frac{C}{16 \pi} \frac{m_\phi^3}{M_{pl}^2}
\end{equation}
where $C$ is a dimensionless constant.  
The branching fraction for modulus decay to gravitino pairs is therefore $C/N$.  

In the approximation of instantaneous modulus decay, the ratio of modulus particles
before the decay to entropy immediately after the decay is simply $Y_\phi = T_R/m_\phi$.
Thus we estimate that
\begin{equation}
Y_{3/2} = \frac{C}{N} \frac{T_R}{m_{\phi}} = 6 \times 10^{-8}  \frac{C}{\sqrt{N}}
\left( \frac{m_\phi}{1000  \ \mathrm{TeV}} \right)^{1/2}
\end{equation}
Now conservation of R-parity implies that each gravitino will eventually decay into at
least one LSP, so we can approximate $Y_{LSP} = Y_{3/2}$.  If the LSPs do not
pair annihilate, then we find
\begin{equation}
\frac{\rho_{LSP}}{s} = 6 \times 10^{-6} \ \mathrm{GeV} \frac{C}{\sqrt{N}}
\left( \frac{m_{LSP}}{100  \ \mathrm{GeV}} \right) \left( \frac{m_\phi}{1000  \ \mathrm{TeV}} \right)^{1/2}
\end{equation}
so we must take $C/\sqrt{N} \lesssim 10^{-4}$ to avoid overclosing the universe.  This 
seems to be a very stringent constraint -- it rules out the most elegant scenario with $C \sim 1$
-- so in the next section we will consider whether a small $C$ can be achieved naturally,
or if other processes can reduce the relic density of LSPs.  Note that the relic density 
is proportional to $m_{LSP}$, so if this scenario can be made viable, then we expect a light spectrum.

One might hope that the LSPs produced by gravitino decay pair annihilate, decreasing their
abundance to within acceptable levels.  We can use equation (\ref{annih}) together with
the spectrum given in equation (\ref{spectrum}) to write the resulting LSP density in terms of $m_{3/2}$
alone \cite{newdecay}, \cite{newdecay2}
\begin{equation} \label{gravannih}
\frac{\rho_{ann}}{s} \approx 8 \times 10^{-9} \ \mathrm{GeV} \frac{(2-x_W)^2}{(1-x_W)^{3/2}}
\left(\frac{m_{3/2}}{100  \ \mathrm{TeV}} \right)^{3/2} ,
\end{equation}
and this would require $m_{3/2} < 20$ TeV to obtain the correct relic density.  Such a light
gravitino would cause severe problems with BBN, and a Wino with $m_{LSP} < 50$ GeV
would already have been detected.  Thus if the gaugino masses are given by anomaly mediation 
(with $\mu \ll m_A$), 
then we cannot rely on Wino pair annihilation, and instead we need a mechanism for suppressing 
$C$ if this scenario is to be viable.  However, because this result seems tantalizingly close to giving
the correct relic density, we will consider below whether deviations from the basic anomaly mediated 
predictions are possible.

\subsection{Possibilities for Suppressing Moduli-Induced Gravitinos}

\subsubsection{Approximate Symmetries}

If the modulus is charged under an approximate symmetry, then its decay rate
to gravitino pairs may be naturally suppressed.  For this to make sense, the 
modulus cannot have a shift symmetry, so it cannot be a volume modulus as in KKLT, 
but it may be a shape modulus.  It is also important that there exists
a sector of light particles with the same symmetry properties, so that modulus decay
to this sector is unsuppressed -- otherwise the overall modulus decay rate would be
small, but the branching fraction to gravitinos would be unchanged.

For instance, as a simple toy model we could consider
\begin{eqnarray}
L & = & \int d^4\theta \left[ X^\dag X + \Phi^\dag \Phi + \Phi_c^\dag \Phi_c + 
\epsilon \frac{\Phi^\dag X^2}{M_{pl}} + \mathrm{H.C.} - \frac{(X^\dag X)^2}{M^2} + ...  \right] \nonumber \\
& & + \int d^2 \theta \left[ \Lambda + \mu^2 X + m_\phi \Phi \Phi_c \right]
\end{eqnarray}
where $\Phi$ and $\Phi_c$ have opposite charge under an approximate $U(1)$ symmetry
whose violation is parameterized by $\epsilon$.  In the limit that $\epsilon \to 0$, the modulus
cannot decay to gravitinos by charge conservation, so for small $\epsilon$ the decay rate is 
suppressed.  We could easily include a sector of light `charged' particles so that the overall
modulus decay rate is not suppressed.  Whether or not such a situation could arise depends
on the details of very high-energy physics, but there is no reason to expect that such a setup
is impossible.

\subsubsection{Light Moduli}

If the modulus is light, with mass close to $2 m_{3/2}$, then the branching fraction to gravitinos
will be phase-space suppressed.  Thus parametrically we would have
\begin{equation}
C \propto \frac{\sqrt{m_{\phi}^2 - 4 m_{3/2}^2}}{m_{\phi}} ,
\end{equation}
but obtaining $C \sim 10^{-3}$ in this way would require a large, poorly motivated fine-tuning.

A more interesting possibility is that there are two or more moduli with different masses.  If the lightest
modulus is lighter than $2m_{3/2}$, then it will not produce gravitinos, but its decay will dilute the
gravitinos from earlier processes.  If the two moduli have masses $m_{\phi} < 2 m_{3/2} \ll m_{\Phi}$,
then the lighter modulus begins rolling later in the universe's history, so that 
$\rho_{\phi} \approx \rho_{\Phi}$ while they are both oscillating.  After $\Phi$ decays, the
lighter field dominates the energy density of the universe until it in turn decays, releasing more entropy.  
Between the time of $\Phi$ decay and the time of $\phi$ decay, radiation red-shifts as $R^{-4}$
but the energy in $\phi$ only red-shifts as $R^{-3}$, so $\phi$ decay dilutes heavy relics such
as gravitinos by a factor
\begin{equation}
\Delta \approx \left( \frac{R_{\phi}}{R_{\Phi}} \right)^{3/4} \approx \left( \frac{T_{RH-\Phi}}{T_{RH-\phi}} \right)^{3/4}
\approx \left( \frac{m_\Phi}{m_\phi} \right)^{9/8}
\end{equation}
which could optimistically be as large as $1000$.  For instance, with $m_\Phi = 5 \times 10^4$ TeV, 
$m_\phi = 100$ TeV, $m_{3/2} = 60$ TeV, and $N \sim 10$, one would need
$C \sim 1/20$ to obtain the correct relic abundance of dark matter (with these values direct production
of dark matter from $\phi$ decay is not a problem).  We will see that the 
presence of such a light modulus is also useful for alleviating cosmological bounds on the axion.

Finally, we could simply return to the original scenario of Moroi and Randall \cite{modtosm}, 
with a single modulus with $m_\phi < 2 m_{3/2}$.  As they showed, the correct
relic density can be obtained rather easily by relying on the pair annihilation studied in section 3.1
(in addition, they claimed that the modulus branching fraction to Winos was chirally 
suppressed, but this is false \cite{newdecay}, \cite{dine}).  It is certainly possible that moduli with masses of order
$m_s$ simply do not exist, but this is a UV sensitive issue.

\subsubsection{Model Dependence}

Although models with $C \sim 1$ are generic, there are models where $C$ is parametrically
different.  For instance, as shown in \cite{dine}, in specific supergravity realizations of the KKLT
scenario, one finds
\begin{equation}
C \sim \left( \frac{m_{3/2}}{m_{\phi}} \right)^2 .
\end{equation}
In this case $F_\phi \sim \frac{m_{3/2}}{m_\phi} F$ as usual, but `$\phi$' is not a mass
eigenstate.  The mass eigenstate modulus mixes with the supersymmetry breaking sector,
so that it effectively has $F_{eigen} \sim \frac{m_{3/2}^2}{m_\phi^2} F$
giving a suppressed $\phi \to 2 \psi_{3/2}$ decay rate.  This result depends on specific properties 
of the supersymmetry breaking sector (the Polonyi field must be light, 
with mass of order $m_{3/2}$) that need not be true in more general KKLT-type scenarios,
but these theories are technically natural, and in some cases the terms that would give
$C \sim 1$ can be forbidden by global symmetries.

We have been assuming that the gauginos get there mass from anomaly mediation alone,
but as mentioned above, it is possible that the same operators
\begin{equation}
\int d^2\theta \frac{\Phi}{M_{pl}} W_\alpha W^\alpha  \ \ \ \implies \ \ \
L \supset \frac{F_\phi}{M_{pl}} \lambda \lambda
\end{equation}
that allow moduli to decay to gauginos also contribute to the mass of the gauginos.
If $F_\phi/M_{pl} \sim m_{3/2}^2/m_\phi$ as expected generically, then the Wino mass can be 
altered, and for a light gravitino and heavy modulus the Wino could become lighter.  
In that case, Wino pair annihilation could be sufficient to achieve the correct relic density;
for instance, with $m_{\tilde{W}} = 70$ GeV and $m_{3/2} = 100$ TeV, one obtains the 
correct abundance.  
Unfortunately this is a UV sensitive question, and it seems to require fine tuning to make
the Wino light.  Furthermore, there is the possibility of introducing CP violation through these 
types of interactions.  Note that these operators are negligable
when $F_\phi$ is small and the $\phi \to 2 \psi_{3/2}$ decay rate is small, which is the
domain of the section 3.3.1.

A better solution is to use a large $\mu$ term in equation (\ref{spectrum}) to make the Wino light.
We would need to abandon the approximate PQ symmetry explaining the top-bottom hierarchy,
but with $\mu \sim m_A$ one obtains a lighter Wino, making it possible to explain the dark matter
relic abundance purely in terms of pair annihilation.

A final possibility is that supersymmetry breaking is dominated by $D$ terms.  As shown in the
appendix, in this case $C$ would be suppressed as
\begin{equation}
C \propto \left( \frac{F}{D} \right)^2 .
\end{equation}
This occurs because only the sgoldstino decays to gravitino pairs, and in the limit of $D$-breaking,
there is no sgoldstino (instead the goldstino has a vector partner, which is not of interest here).
As shown in \cite{Dterms} it is possible to obtain parametrically large $D$ terms, and $D$-breaking
naturally gives a split supersymmetry type spectrum. This possibility may warrant further investigation,
as it may help to alleviate the cosmological moduli/gravitino problem in more general circumstances.

\subsection{The $\phi \to \tilde{\phi} + \psi_{3/2}$ Channel}

For completeness, we consider the decay channel $\phi \to \tilde{\phi} + \psi_{3/2}$ 
\cite{modtogravmod}, although we will see that it is dangerous, so we will need to forbid 
it kinematically.  It comes from operators of the form
\begin{eqnarray}
L & \supset &  \int d^4 \theta \frac{a}{M_{pl}} X \Phi^\dag \Phi \nonumber \\
& = & a \frac{m_{\tilde{\phi}}}{M_{pl}} \chi \tilde{\phi} \phi -
\sqrt{3} a m_{3/2} m_{\tilde{\phi}} \phi \phi + ...
\end{eqnarray}
where $\chi$ is the goldstino, $X = x + \theta \chi + \theta^2 m_{3/2} M_{pl}$ is the field that
dominantly breaks supersymmetry, and $a$ is an $O(1)$ coupling constant that must be relatively large 
so that the modulino will be lighter than the modulus.  
As a complex scalar field, $\phi$ actually has two real modes, with masses
\begin{equation}
m_{\phi \pm} = \sqrt{m_{\tilde{\phi}}^2 \pm \sqrt{3} a m_{3/2} m_{\tilde{\phi}}}
\approx m_{\tilde{\phi}} \pm \frac{\sqrt{3}}{2} a m_{3/2}
\end{equation}
Generically, both modes will be present, but $\phi_-$ cannot decay to a modulino and a 
gravitino.  If we require that 
$m_{\phi +}$ is greater than the sum of the modulino and gravitino masses, then we must 
have $a > (2 + m_{3/2}/m_{\tilde{\phi}})/\sqrt{3}$; if this relation is violated, then
the decay will be kinematically prohibited.

With these assumptions, one can calculate the modulus decay rate and branching 
fraction into modulinos and gravitinos \cite{modtogravmod}, obtaining
\begin{eqnarray}
\Gamma & = & \frac{a^2 m_\phi m_{\tilde{\phi}}^2}{8 \pi M_{pl}^2}
\left(1 -\frac{(m_{\tilde{\phi}} + m_{3/2})^2}{m_\phi^2}\right)^{3/2}
\left(1 -\frac{(m_{\tilde{\phi}} - m_{3/2})^2}{m_\phi^2} \right)^{1/2} \nonumber \\
& \approx & \frac{3 \sqrt{2} a^4 m_{\phi} m_{3/2}^2}{8 \pi M_{pl}^2}
\end{eqnarray}
for the decay rate into a gravitino and a modulino, where in the second line we take $m_{3/2} \ll m_{\phi}$.
Thus we find a relic abundance \cite{modtogravmod}
\begin{equation}
Y_{3/2}  \gtrsim 10^{-9} \left(\frac{m_{3/2}}{100 \ \mathrm{TeV}}\right)^2 
\left(\frac{1000 \ \mathrm{TeV}}{m_{\phi}}\right)^{3/2} 
\end{equation}
We know that $Y_{LSP} \approx Y_{3/2}$, so we obtain 
\begin{equation}
m_{LSP} Y_{LSP} \gtrsim 10^{-7} \ \mathrm{GeV} \times \left(\frac{m_{LSP}}{100 \ \mathrm{GeV}}\right) 
\left(\frac{m_{3/2}}{100 \ \mathrm{TeV}}\right)^2 \left(\frac{1000 \ \mathrm{TeV}}{m_{\phi}}\right)^{3/2} 
\end{equation}
for the dark matter energy density of the universe. 
We expect that $m_{LSP} Y_{LSP} = 4 \times 10^{-10}$ GeV \cite{WMAP} if the LSP
accounts for all of the dark matter in the universe, so this decay mode will 
overproduce LSPs.  Fortunately, we can prohibit this decay kinematically 
without introducing any fine-tuning.

\subsection{Weakening Cosmological Bounds on the Axion}

The axion \cite{origaxion} decay constant $F$ is bounded from below due to astrophysical 
constraints, and in generic cosmological scenarios it is bounded from above by the requirement 
that the axion does not overclose the universe.  The moduli fields in our model
decay when the universe is at a temperature near $\Lambda_{QCD}$, releasing a significant
amount of entropy, and potentially relaxing constraints on $F$ by diluting the axions.  Throughout
we will be considering the relic axion energy density from misalignment production, which is by
far the dominant method of production for large $F$.

As shown in \cite{axionblow}, diluting relic axions is a bit delicate.  This is because there
are three processes that need to be considered:  a particle or field is decaying, releasing entropy
and making the universe cool more slowly; $H$ is decreasing, alleviating axion hubble friction;
and the axion mass is increasing, since $m_a$ is strongly dependent on temperature because it arises
from instanton effects.  The axion only begins to roll as $H$ drops below $m_a$, so if our modulus
decays before this point then the entropy released does not decrease the final axion density.  In 
general, particles decaying after the universe has cooled below $1$ MeV are dangerous to BBN, 
leaving a narrow window of $1$ MeV $< T_{R} < T_h$ for reheating temperatures 
of decays that can dilute axions.  Note that $T_R$ should be interpreted as the approximate 
temperature of the universe after modulus decay -- the temperature of the universe never increases.

First we estimate $T_h$, the reheating temperature above which the modulus does not dilute the
axion density at all.  As the modulus decay completes, the universe will be radiation dominated 
with temperature $T_R$, so $H \sim T_R^2/M_{pl}$ at this time.  The axion has mass
\begin{equation}
m_a(0) = 13 \ \mathrm{MeV} \left( \frac{1 \ \mathrm{GeV}}{F} \right)
\end{equation}
at temperatures less than about $\Lambda_{QCD}/\pi$ and
\begin{equation}
m_a(T) = 0.1 m_a(0) \left( \frac{\Lambda_{QCD}}{T} \right)^{3.7}.
\end{equation}
at temperatures above $\Lambda_{QCD}/\pi$.  If $T_R = T_h$, then $3H(T_h) = m_a(T_h)$ 
just as the modulus decays, so we find that
\begin{equation}
T_h = 1.5 \ \mathrm{GeV} \left(\frac{10^{12} \ \mathrm{GeV}}{F} \right)^{0.18}
\end{equation}
where we have taken $\Lambda_{QCD} = 200$ MeV.  We see that modulus decay will certainly
dilute axions with $F$ near the current bound, but for $F = 10^{15}$ GeV, $T_h \sim T_R$
for the $1000$ TeV modulus field.  However, we can do much better with a lighter modulus.

In order to obtain a relic axion energy density less than the current matter density in the universe, it
was shown in \cite{axionblow} that we must have
\begin{equation}
F < 2 \times 10^{14} \ \mathrm{GeV} \left(\frac{100 \ \mathrm{MeV}}{T_R} \right)^{1/2} 
\left( \frac{m_a(T_a)}{m_a(0)} \right)^{1/2}
\end{equation}
where $T_a$ is the temperature of the universe when $m_a(T) = 3H$.
We can estimate $T_a$ as \cite{axionblow} 
\begin{equation}
T_a = 200 \ \mathrm{MeV} \left(\frac{T_R}{100 \ \mathrm{MeV}}\right)^{0.26} \left(\frac{10^{15} \ \mathrm{GeV}}{F}\right)^{0.13}
\end{equation}
thus as anticipated above, $m_\phi = 1000$ TeV leads to no improvement.
However, if we consider moduli with $m_\phi \sim 100$ TeV as considered in section 
3.1 and 3.3.2 then we obtain $T_R \sim 6$ MeV, giving $F < 7 \times 10^{14}$ GeV.  

\section{Displaced Gluino Vertices at the LHC}

We know that cosmology favors a light anomaly mediated spectrum, so we can 
be justifiably optimistic that our model will be tested at the LHC.  In fact, we expect that
gluinos will be copiously produced, and that gluino pairs will decay at secondary
vertices separated by distances of order a millimeter or more.  To begin to analyze this 
process, note that the resolution of the LHC's ATLAS detector \cite{ATLAS} will be
\begin{eqnarray}
\Delta d_0 & = & 11 + \frac{73}{(p_T / \mathrm{GeV}) \sqrt{\sin(\theta)}} \ \mu m \\
\Delta z_0 & = & 87 + \frac{115}{(p_T / \mathrm{GeV}) (\sin(\theta))^{\frac{3}{2}}} \ \mu m
\end{eqnarray}
Since gluinos are produced by QCD reactions such as $gg \to \tilde{g} \tilde{g}$, we can
expect a large $p_T$ and an order one $\theta$.
Thus we will need gluino decay vertices displaced by $\gtrsim 10$ $\mu m$ to have a chance of
distinguishing them from primary vertices.  There are four factors relevant to the displacement
distance:  the gluino production rate, the gluino lifetime, the relativistic time dilation, 
and the possibility of detecting gluinos from the tail of their distribution, 
which decays exponentially with distance from the primary vertex.

We will not attempt to compute the the gluino production rate precisely \cite{gluinoprod}, \cite{gluinoprod2}, 
\cite{gluinoprod3}, \cite{gluinoprod4}, but we know that 
the answer is large, and our results will be very insensitive to the details.  It was found in \cite{stopgluino}
that for $m_{\tilde{g}} \sim 350$ GeV, the LHC will produce about one gluino per second, 
or about $3 \times 10^7$ per year of operation.  For $m_{\tilde{g}} \sim 2$ TeV, the LHC 
will produce at least a thousand per year.  Thus we make the conservative assumption that the 
LHC will produce
\begin{equation}
N = 10^3 \left(\frac{2 \ \mathrm{TeV}}{m_{\tilde{g}}}\right)^4
\end{equation}
gluinos per year.

Various groups \cite{gluinodecay}, \cite{gluinochannels} have examined the decay of the gluino
and computed its lifetime in Split Supersymmetry, 
including the relevant one-loop operator running from the SUSY breaking scale to 
the TeV scale.  The result is
\begin{equation}
\tau_{g} = \frac{4 \ \mathrm{sec}}{K} \left(\frac{m_s}{10^6 \ \mathrm{TeV}}\right)^4
\left(\frac{1 \ \mathrm{TeV}}{m_{\tilde{g}}}\right)^5
\end{equation}
where $K \lesssim 1$ is a weakly varying function of $m_s$, $m_{\tilde{g}}$, 
and $\tan(\beta)$.  This is extended by a factor of approximately
\begin{equation}
\gamma \approx \frac{3 \ \mathrm{TeV}}{m_{\tilde{g}}}
\end{equation}
due to relativistic time dilation.  We translate the gluino lifetime into a displacement by 
multiplying by a factor of $c \sqrt{1-m^2/(3 \ \mathrm{TeV})^2} \approx c$.

\begin{figure}[th]
\begin{center}
\includegraphics[width=14cm, height=8cm]{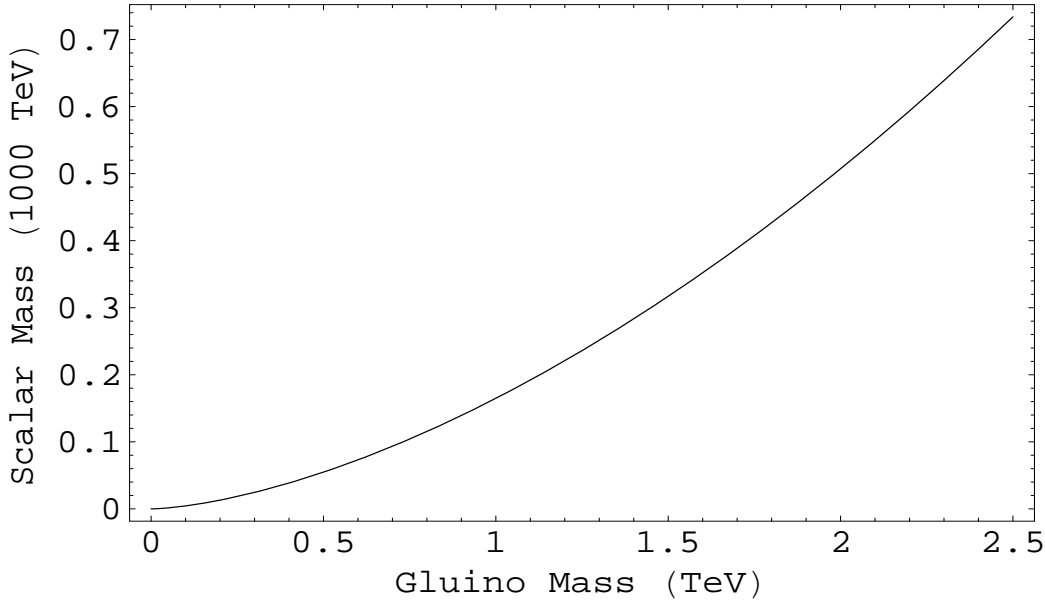}
\caption{\small{We will observe displaced gluino vertices at the LHC if $m_s$ and 
$m_{\tilde{g}}$ lie in the region above the curve.}}
\end{center}
\end{figure}

Putting these results together, we find that the number of gluinos with 
displacement in the interval $(R, R + dR)$ is
\begin{equation}
N(R) = 1000 \left( \frac{2 \ \mathrm{TeV}}{m_{\tilde{g}}} \right)^4
\exp\left(-\frac{R}{D_0}\right) \frac{dR}{D_0}
\end{equation}
where the canonical distance $D_0$ is given by
\begin{equation}
D_0 = \left(7.2 \  \mathrm{mm}\right) \left(\frac{1 \ \mathrm{TeV}}{m_{\tilde{g}}}\right)^6
\left(\frac{m_s}{1000 \ \mathrm{TeV}}\right)^4
\end{equation}
If we require the observation of at least a few displaced gluino vertices in one year's worth of
LHC data, we find the limit
\begin{equation}
10^{-2} < \left(7 + 4 \log\left( \frac{2 \ \mathrm{TeV}}{m_{\tilde{g}}} \right) \right)
\left(\frac{1 \ \mathrm{TeV}}{m_{\tilde{g}}}\right)^6
\left(\frac{m_s}{1000 \ \mathrm{TeV}}\right)^4
\end{equation}
on the mass parameters of the model.  This result is only accurate to within a factor of a few, 
but this is unimportant because displaced gluino vertices will be visible for virtually the entirety 
of our anomaly-mediated parameter space, as seen in figure 1.  For comparison, if we require
the observation of $100$ displaced gluino vertices each year at the LHC, then the digit
$7$ in the above equation changes into a $2$, but this is the only alteration.  Note that 
backgrounds for this signal from muons and b-quarks will be 
small due to the large jet energies, large missing energy, and especially because of the large 
vertex displacements.  It was shown in the detailed study of \cite{RHadron} that simply
using cuts on $E_T$ and missing $E_T$, for $m_{\tilde{g}} < 1.4$ TeV, the signal from 
R-Hadronized gluinos could be seen in $30$ $fb^{-1}$ of LHC data.

\section{Conclusions}

We have shown that a simple Split Supersymmetry spectrum based on
anomaly mediation can satisfy cosmological constraints with a light spectrum.  
This is possible because the successive decays moduli $\to$ LSPs and moduli $\to$ gravitinos 
$\to$ LSPs generate dark matter very efficiently, in contrast with most anthropically motived models 
\cite{aspsplit}, \cite{wells}, \cite{mindmunif} based on perturbative freeze out, 
which often require $1$-$2$ TeV LSP masses (\cite{PTDM} is an exception).  Unfortunately,
in one of the most elegant scenarios dark matter is overproduced, but we found several
mechanisms that can lead to the correct relic density, and the original mechanism of Randall
and Moroi \cite{modtosm} remains a very viable possibility.  We have also seen that our model
will have a striking LHC signature -- displaced gluino vertices.

As outlined in the introduction, there are only a few clues for would-be anthropic model
builders.  The two most tantalizing are probably gauge coupling unification and dark matter,
and we must assume that these two issues are resolved in concert if we are to avoid
an almost infinite set of possibilities for new physics.
Here we have shown that retracting the usual assumptions about the generation of dark matter do
not necessarily make models irrelevant for experimental collider physics, as we might have
feared\footnote{However, it is worth noting that when dark matter is not generated by perturbative
freeze out, we lose the elegant parametric prediction that the weak scale is the geometric mean of the Planck
and Cosmological Constant scales \cite{split}.  The numerical prediction for the value of the cosmological constant
is unchanged.}.  Furthermore, we noticed a fortuitous accident -- as a side effect, our mechanism
weakens cosmological constraints on axions, expanding the rather narrow window on the
axion decay constant.  With so little information about new physics, we should be appreciative
when a small piece of the strong CP problem falls into our lap.

The success of anthropic predictions 
of the cosmological constant, combined with the derth of electric dipole moment signatures 
and new flavor changing neutral current effects, the smallness of the S and T parameters, 
and the lack of new physics at LEP point toward an anthropic solution to the two 
naturalness problems of high-energy physics and cosmology.  If the world is supersymmetric
at high energy, then the spectrum of new particles that we have considered is an 
excitingly predictive model for LHC physics.

\appendix
\section{Appendix: Moduli-Gravitino Decay Rates}

These results were obtained in collaboration with Aaron Pierce and Jesse Thaler.  Similar methods can
be found in \cite{lutygoldstino}.

Consider an arbitrary, globally supersymmetric lagrangian $L(\phi, \psi, F, \lambda, A_\mu, D, \partial_\mu)$
that is a function of off-shell chiral multiplets and vector multiplets.  Now assume that
supersymmetry is spontaneously broken by some combination of $D$ component 
and $F$ component VEVs.  
The sgoldstino\footnote{If there are non-vanishing $D$ component
VEVs then there will also be a vector partner of the goldstino, but since we are only 
interested in moduli-goldstino interactions, we are ignoring these fields.} is the linear combination
\begin{equation}
\phi_\chi = \sum_i \frac{F_i}{F} \phi_i ,
\end{equation}
where $F^2 = \Sigma_i F_i^2 + \Sigma_a D_a^2$ is the goldstino decay constant.  Note
that there is no reason to expect that $\phi_\chi$ is a mass eigenstate.
We will show that the only two derivative operator involving a scalar and two goldstinos is
\begin{equation}
\frac{1}{F^\dag} \partial_\nu \phi_\chi^\dag \chi \sigma^\nu \bar{\sigma}^\mu \partial_\mu \chi 
- \frac{1}{F} \partial_\nu \phi_\chi \chi^\dag \bar{\sigma}^\mu \sigma^\nu \chi^\dag .
\end{equation}
Thus the rate of moduli decays to gravitino pairs is determined by the overlap of 
the mass-eigenstate moduli with $\phi_\chi$.  As a simple corollary, we see that if
supersymmetry were broken entirely by $D$ terms, then there would not be a sgoldstino, 
so the rate for modulus decay to gravitino pairs would be zero.

To prove this assertion, we need to do a field redefinition to introduce the non-linearly
realized goldstino $\chi$.  As in the conceptually simpler case of a goldstone boson,
we can do this by choosing a `vacuum alignment', and then parameterizing
a supersymmetry transformation with the goldstino $\chi(x)$.
For instance, if the theory consisted of a single chiral multiplet $\Phi = (\phi, \psi, F)$,
we would represent $\Phi = e^{i \delta_\chi} (\phi, 0, F)$.

In the general case, when supersymmetry is linearly realized the goldstino is the combination
\begin{equation}
\psi_g = \sum_i \frac{F_i}{F} \psi_i + \sum_a \frac{D_a}{F} \lambda_a .
\end{equation}
To isolate it, we begin by rotating the fermions $\psi_i$, $\lambda_a$ 
into a mass eigenstate basis, so that the fermion lagrangian becomes
\begin{equation}
L_f = i \psi_g^\dag \bar{\sigma}^\mu \partial_\mu \psi_g
+ \sum_\alpha \left[ i \psi_{\alpha}^\dag \bar{\sigma}^\mu \partial_\mu \psi_{\alpha} 
+ \frac{m_\alpha}{2} ( \psi_{\alpha} \psi_{\alpha} + \psi_{\alpha}^\dag \psi_{\alpha}^\dag) \right] + ...
\end{equation}
where the elipsis denotes interaction terms.  Note that the superpartners of the 
fermions $\psi_\alpha$ have no auxiliary component VEVs.

Now we perform a $\chi$ parameterized supersymmetry transformation on $L$,
and then choose the `vacuum alignment' condition $\psi_g = 0$.  We are only interested
in two operators, the interaction term mentioned above and the kinetic term for $\chi$, which we require
for canonical normalization.  Both of these operators involve two goldstinos, so they can only
come from $C \delta_\chi^2 A$ or $C \delta_\chi A \delta_\chi B$ where $A$, $B$, and $C$ 
are some combination of fields in the lagrangian.  But the first type of term would be
$\frac{\Delta L}{\Delta A} \delta_\chi^2 A$, and this vanishes on-shell because it is proportional to the $A$ 
equation of motion.  Thus only terms of the second kind are relevant.
 
The supersymmetry transformation rules are the familiar
\begin{eqnarray}
\delta_\chi \phi & = & \sqrt{2} \chi \psi   \\
\delta_\chi \psi & = & i \sqrt{2} \sigma^\mu \chi^\dag D_\mu \phi + \sqrt{2} \chi F \\
\delta_\chi F & = & i \sqrt{2} \chi^\dag \bar{\sigma}^\mu D_\mu \psi \\
\delta_\chi \lambda & = & \sigma^{\mu \nu} \chi F_{\mu \nu} + i \chi D \\
\delta_\chi A_\mu & = & -i \lambda^\dag \bar{\sigma}_\mu \chi + i \chi^\dag \bar{\sigma}_\mu \lambda \\
\delta_\chi D & = & - \chi \sigma^\mu D_\mu \lambda^\dag - D_\mu \lambda \sigma^\mu \chi^\dag .
\end{eqnarray}
where $D_\mu$ is the appropriate gauge covariant derivative.
The $\chi$ kinetic term can only come from the transformation of the $\psi_g$ kinetic term,
since the transformation of all other operators involve too many fields.  Furthermore, it is
remarkable that two derivative interactions between a scalar and two goldstinos also
only come from the $\psi_g$ kinetic term, because the other terms either give too many
fields or vanish on the equations of motion.  Thus as claimed we find
\begin{equation}
L \supset i \chi^\dag \bar{\sigma}^\mu \partial_\mu \chi 
+ \frac{1}{F^\dag} \partial_\nu \phi_\chi^\dag \chi \sigma^\nu \bar{\sigma}^\mu \partial_\mu \chi 
- \frac{1}{F} \partial_\nu \phi_\chi \chi^\dag \bar{\sigma}^\mu \sigma^\nu \chi^\dag + ...
\end{equation}
after canonically normalizing $\chi$, where the elipsis denotes other interactions, including a
two goldstino interaction with the vector superpartner of the goldstino.  Although from
the transformation rules it might seem that $\chi$ has non-derivative interactions, these all
cancel -- $\chi$ has a shift symmetry, just like a goldstone boson.  Thus yukawa couplings
with $\chi$ cannot be generated.

We should note that this is an effective field theory, and it is only
valid for $E^2 \ll F = m_{3/2} M_{pl}$.  In most examples of interest (KKLT moduli, 
or moduli that get masses from supersymmetry breaking), the modulus mass
easily satisfies this criterion.  In any case, above this scale supersymmetry breaking is 
a small effect, and in place of a single goldstino one would need to consider the detailed
dynamics of the hidden sector.

\section*{Acknowledgements}

The results of Appendix A were obtained in collaboration with Aaron Pierce and Jesse Thaler.
We would like to thank Lian-Tao Wang, Ben Lillie, Philip Schuster, Natalia Toro, and especially 
Nima Arkani-Hamed for discussions.  This work was supported by the Fannie and John Hertz Foundation.

\end{document}